\documentclass{article}
\usepackage{arxiv}

\usepackage[T1]{fontenc}
\usepackage[utf8]{inputenc}
\usepackage{amsmath,amssymb}
\usepackage{graphicx}
\usepackage{booktabs}
\usepackage{tabularx}
\usepackage{flafter}
\usepackage[numbers,sort&compress]{natbib}
\usepackage{tikz}
\usetikzlibrary{arrows.meta,positioning,fit,backgrounds,calc}
\usepackage{listings}
\usepackage{xcolor}
\usepackage[colorlinks=true,linkcolor=blue!55!black,citecolor=blue!55!black,urlcolor=blue!55!black]{hyperref}
\providecommand{\doi}[1]{\href{https://doi.org/#1}{doi:#1}}
\usepackage[capitalise]{cleveref}

\newcommand{\cadaques}{\textsc{cadaques}}
\newcommand{\Tc}{T_{\mathrm{c}}}
\newcommand{\code}[1]{\texttt{#1}}

\renewcommand{\headeright}{Preprint}
\renewcommand{\undertitle}{Preprint}

\title{\cadaques{}: A Cost-Aware Dual Architecture for\\ Query-Efficient Autonomous Discovery}

\author{Jorge Bravo-Abad \\
  Departamento de F\'isica Te\'orica de la Materia Condensada and\\
  Condensed Matter Physics Center (IFIMAC)\\
  Universidad Aut\'onoma de Madrid, E-28049 Madrid, Spain\\
  \texttt{jorge.bravo@uam.es}
}

\hypersetup{
  pdftitle={CADAQUES: A Cost-Aware Dual Architecture for Query-Efficient Autonomous Discovery},
  pdfauthor={Jorge Bravo-Abad},
  pdfsubject={Cost-aware software architecture for autonomous discovery campaigns},
  pdfkeywords={autonomous discovery, cost-aware optimization, scientific software, provenance, Bayesian optimization}
}

\begin{document}
\maketitle

\begin{abstract}
Autonomous discovery systems couple a resource that answers queries (a simulator,
instrument, or analytic model) to an algorithm that selects
what to query next. Most software frameworks for this loop inherit
the control structure of numerical optimization: campaigns run for a fixed number
of iterations, query costs are not represented in the programming interface, and
the cost of decision-making is treated as negligible. In practice, queries may differ in cost by
orders of magnitude, and planners built on large language
models or expensive surrogates consume resources of their own. Here we present
\cadaques{}, an open-source Python framework built on one architectural principle: cost is a first-class primitive of the discovery loop. \cadaques{}
separates the loop into two structural protocols, an \emph{Oracle} that answers
queries and a \emph{Driver} that proposes them, and charges oracle
evaluations and driver decisions against a common vector-valued budget spanning
wall time, CPU hours, monetary cost, and language model tokens.
For each transaction, an append-only ledger records the cost declared before
execution and the cost settled afterwards, making their discrepancy an observable
property of the campaign. We evaluate its architectural claims by locating the
critical temperature of the two-dimensional Ising model from noisy finite-size
estimates against an exact thermodynamic-limit reference. In this deliberately
noisy setting, strategies that concentrate around the best observed result can be
misled by peaks produced by noise, whereas a schedule that explores with low-cost, low-fidelity queries and
refines with higher-fidelity ones produces, in this experiment and at the studied budget scale, lower and less variable observed errors than using high
fidelity throughout. Metering adds tens
of microseconds per iteration, three orders of magnitude below the
cheapest oracle query. The framework is MIT-licensed and archived at
Zenodo (\doi{10.5281/zenodo.21293589}).
\end{abstract}


\section{Introduction}
\label{sec:intro}

Closed-loop autonomous discovery is increasingly used across the natural sciences.
Self-driving laboratories integrate synthesis, characterization, and experiment
planning in chemistry and materials science~\cite{Abolhasani2023,MacLeod2020,Roch2018}.
Bayesian optimization provides widely used machinery for sample-efficient experiment
selection~\cite{Shahriari2016,Snoek2012,Balandat2020}. More recently, agents based on
large language models (LLMs) have begun to plan and execute multi-step scientific
workflows~\cite{Boiko2023,Bran2024}. Despite the diversity of these settings, they
share the same computational abstraction: an expensive resource answers queries,
and an algorithm decides what to query next.

Software for implementing this loop generally inherits its control structure from
numerical optimization, along with three assumptions that are increasingly difficult
to justify in autonomous discovery. First, campaigns are often governed by
a fixed number of evaluations and assessed by the best value obtained after \(N\)
iterations. This convention implicitly assigns equal cost to every query, even though nominally
similar queries may differ in cost by orders of magnitude. For
instance, in materials science, the same candidate atomic structure may be evaluated
using either a tightly converged density-functional calculation or a much cheaper
force-field model, with substantial differences in computational cost and fidelity.

Second, query cost is rarely represented explicitly in the interface between the
optimizer and the objective. Cost-aware methods, including multi-fidelity
strategies~\cite{Kandasamy2017}, must therefore rely on framework-specific extensions
or conventions outside the main campaign abstraction. This makes it difficult to
express strategies that adapt their behavior to the remaining budget or choose the
fidelity of each query according to its expected value and cost. Third, the
decision-making procedure itself is commonly treated as free. That approximation may
be acceptable for inexpensive heuristics, but it becomes less plausible for planners
that repeatedly fit Gaussian process surrogates or invoke LLMs. In such cases,
deliberation consumes wall time, compute, tokens, or funds, and may account for a
non-negligible fraction of the total campaign expenditure.

These assumptions have several practical consequences. Comparisons based only on iteration
counts are difficult to interpret when per-query costs differ. Resource expenditure
must often be reconstructed after the fact from heterogeneous logs, when it is
recorded at all. Budget-adaptive strategies are awkward to implement when neither
query cost nor remaining resources are exposed through the programming interface.
More generally, they make it difficult to determine whether an expensive planner
justifies the resources it consumes.

This paper presents \cadaques{} (Cost-Aware Dual Architecture for QUery-Efficient autonomous diScovery), a compact open-source Python
framework that treats cost as a first-class primitive of the discovery loop. Rather
than adding cost tracking to an iteration-driven optimizer, \cadaques{} represents
each expense as a typed, vector-valued resource cost and charges it against a campaign
budget whose capped components determine termination.

The framework separates the discovery loop into two structural protocols. An
\emph{Oracle} answers queries through the methods \texttt{price} and
\texttt{evaluate}, while a \emph{Driver} selects queries through
\texttt{propose} and \texttt{observe}. Structural typing allows numerical simulators, laboratory instruments, classical
optimizers, and language model agents to occupy their respective roles
without inheriting from framework-specific base classes. Both oracle evaluations and
driver decisions are charged against the same vector-valued budget, with separate
components for resources such as wall time, CPU hours, monetary cost, and
language model tokens. For every transaction, \cadaques{} distinguishes between the
cost declared before execution and the cost settled afterwards. The declared cost
supports affordability checks and planning, while the settled cost records what was
actually spent. Their difference becomes an observable property of the campaign
rather than an error hidden in external logs. An append-only ledger stores these
records in a machine-readable form and supports the reconstruction of resource-normalized
performance measures directly from the transaction history.

The paper makes three contributions. First, it reduces the integration surface of
an autonomous discovery campaign to two role-specific interfaces: an \emph{Oracle}
(\texttt{price}, \texttt{evaluate}) and a \emph{Driver} (\texttt{propose},
\texttt{observe}). Second, it introduces vector-valued budgets, separate declared and
settled costs, and common accounting for queries and decisions. Third, it implements
these ideas in a dependency-light Python package with a campaign runner, an
append-only ledger, and reference Oracles and Drivers, distributed through PyPI and
archived at Zenodo.

We test the design on a controlled physics task: locating the critical region of the
two-dimensional Ising model from noisy finite-size susceptibility estimates. The
thermodynamic-limit critical temperature provides an external reference. The study
compares three protocol-conformant Drivers under equal nominal wall-time budgets,
includes the proposal cost of a Gaussian process planner, examines fixed and scheduled
fidelity policies, and measures the overhead of metering and ledger updates.

The experiments expose several practical effects. Strategies that concentrate around
the best observation can be trapped by noise peaks. A simple schedule that uses cheap,
low-fidelity queries for exploration and more expensive queries for refinement produces,
in this experiment and at the studied budget scale, lower and less variable observed
errors than using high fidelity throughout. Metering adds only
tens of microseconds per iteration, roughly three orders of magnitude below the
runtime of the cheapest Oracle query in the study.

The remainder of the paper is organized as follows. Section~\ref{sec:background}
develops the requirements that motivate the design. Section~\ref{sec:design}
presents the architecture, and Section~\ref{sec:impl} describes its
implementation. Section~\ref{sec:eval} reports the evaluation,
Section~\ref{sec:related} reviews related work, Section~\ref{sec:discussion}
discusses limitations and future directions, and Section~\ref{sec:conclusion}
concludes.

\section{Background and Requirements}
\label{sec:background}

The case for treating cost as an architectural primitive follows from how discovery
campaigns consume resources in practice. This section identifies three recurring
cost regimes that are handled poorly by conventional iteration-budgeted loops and
derives from them the five requirements that motivate the design presented in
\cref{sec:design}.

\subsection{Cost regimes in autonomous discovery}

Three recurring regimes illustrate why cost belongs in a discovery framework's
architecture rather than only in user-written post-processing scripts.

The first regime is heterogeneous query cost. In simulation-driven discovery,
fidelity parameters such as mesh density, basis-set size, and statistical sampling
effort trade computational cost against properties such as accuracy, variance, and
resolution. Laboratory analogues include integration time, measurement resolution,
and sample-preparation effort. A campaign that selects the fidelity of each query
therefore operates in a setting where the cost of evaluating the same design point
may vary by orders of magnitude. Multi-fidelity strategies are designed precisely
to exploit this structure~\cite{Kandasamy2017}. An interface that does not represent
per-query cost and fidelity cannot express this trade-off directly, while an
iteration budget obscures it by assigning the same unit weight to evaluations of
very different expense.

The second regime is the coexistence of heterogeneous, only partially commensurable
resources. Laboratory campaigns may consume wall time,
instrument time, consumables, and funds, while computational campaigns consume
wall time, CPU or GPU hours, and, increasingly, metered API calls. These
resources can be combined only through a valuation supplied for a particular
application: there is no universal or stable exchange rate between, for example, an
hour of beamline access and an hour of CPU time. Their relative value depends on the
facility, scientific objective, deadline, and stage of the project. Collapsing them
into a single scalar inside the framework would therefore impose a valuation that
may not transfer across campaigns. Recording expenditure separately by resource
instead supports transparent reporting and allows users to apply their own
valuations at analysis time.

The third regime is costly deliberation. A planner based on standard
exact Gaussian process inference may incur computational costs that grow cubically
with the number of observations, while a planner implemented as an LLM agent may
consume tokens billed at externally metered rates~\cite{Boiko2023}. Treating such
decision-making as free introduces a systematic omission from campaign accounting.
Depending on the planner and oracle, deliberation cost may be negligible, comparable
to the cost of the queries, or even dominant. Whether an expensive but
sample-efficient planner outperforms a cheaper, more sample-intensive alternative
under common resource constraints is therefore an empirical question. It cannot be
answered cleanly by a framework that meters the oracle but not the planner.

\subsection{Requirements}
\label{sec:requirements}

These regimes motivate five requirements for a cost-aware discovery framework.

\begin{itemize}
\item[\textbf{R1}] \emph{Typed, vector-valued resource cost.} Cost must be represented as a
structured value over heterogeneous resource components and attached to every
metered query and decision.

\item[\textbf{R2}] \emph{Budget-governed admission and termination.} The runner must reject a
transaction whose declared cost does not fit within the remaining capped resources
and terminate when the next proposed transaction cannot be admitted under the remaining capped resources. Because
realized cost is known only after execution, settled expenditure may exceed a
nominal cap when a declaration underestimates the actual cost. Comparisons under
common resource constraints are therefore governed by the execution system rather
than reconstructed afterwards by the analyst.

\item[\textbf{R3}] \emph{Declared and settled costs.} The framework must distinguish
the cost estimated before a transaction, which is needed for affordability checks
and planning, from the cost actually incurred, which is needed for accurate
accounting. Both values and their discrepancy must be recorded.

\item[\textbf{R4}] \emph{Symmetric charging.} Oracle evaluations and driver decisions
must be charged against the same campaign budget. The runner should measure
resources it can observe directly, such as wall time, while instrumented
adapters may report additional components such as tokens, compute hours or monetary
charges. Wall time is measured by the runner, whereas CPU hours represent aggregate
compute consumption and must be supplied by an adapter or external scheduler when
they differ from elapsed time.

\item[\textbf{R5}] \emph{Transactional provenance.} The complete sequence of metered
transactions must be exportable in a machine-readable form containing enough
information to reconstruct cumulative resource expenditure and the resource-normalized
performance measures supported by the recorded data.
\end{itemize}

Within the documented public interfaces examined for this work, we did not identify a
framework for autonomous discovery that combines R1--R5 in one protocol-level campaign
abstraction; \cref{sec:related} gives the scoped comparison. The individual mechanisms
have clear precedents. \cadaques{} contributes a compact interface that brings them
together.

\section{Design}
\label{sec:design}

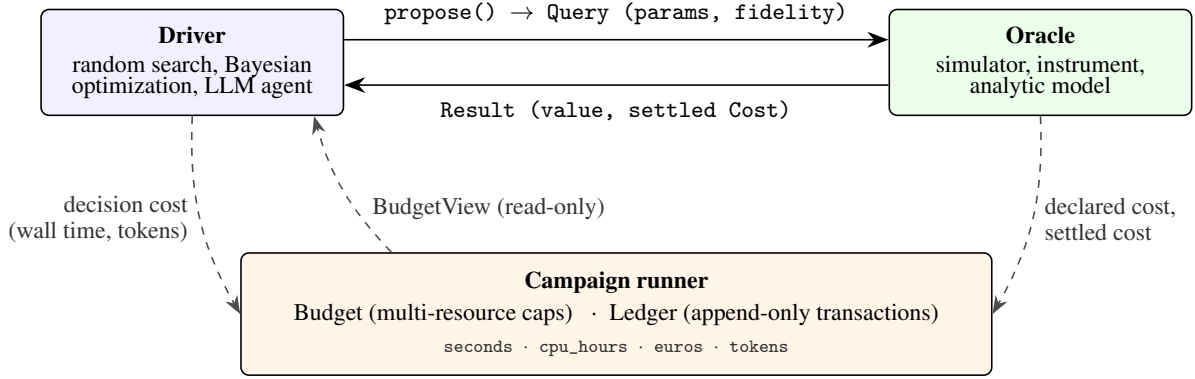
\begin{figure}[t]
\centering
\begin{tikzpicture}[
  font=\small,
  role/.style={draw, rounded corners=3pt, minimum height=14mm, minimum width=40mm,
               align=center, semithick},
  arr/.style={-{Stealth[length=2.8mm]}, semithick},
  marr/.style={-{Stealth[length=2.4mm]}, semithick, dashed, gray!60!black},
  lab/.style={font=\footnotesize\ttfamily},
  mlab/.style={font=\footnotesize, gray!40!black},
]
\node[role, fill=blue!6] (driver)
  {\textbf{Driver}\\[1pt]\footnotesize random search, Bayesian\\[-2pt]
   \footnotesize optimization, LLM agent};
\node[role, fill=green!8, right=72mm of driver] (oracle)
  {\textbf{Oracle}\\[1pt]\footnotesize simulator, instrument,\\[-2pt]
   \footnotesize analytic model};

\draw[arr] ([yshift=3mm]driver.east) --
  node[above=2pt, lab]{propose() $\rightarrow$ Query (params, fidelity)}
  ([yshift=3mm]oracle.west);
\draw[arr] ([yshift=-3mm]oracle.west) --
  node[below=2pt, lab]{Result (value, settled Cost)}
  ([yshift=-3mm]driver.east);

\node[draw, rounded corners=3pt, fill=orange!9, minimum height=16mm,
      inner xsep=7mm, inner ysep=2.5mm,
      align=center, semithick, below=18mm of $(driver.south)!0.5!(oracle.south)$]
  (runner)
  {\textbf{Campaign runner}\\[2pt]
   \footnotesize Budget (multi-resource caps) \enspace$\cdot$\enspace
   Ledger (append-only transactions)\\[2pt]
   \scriptsize\ttfamily\color{gray!40!black} seconds $\cdot$ cpu\_hours $\cdot$
   euros $\cdot$ tokens};

\draw[marr] (driver.south) to[bend right=16]
  node[left=1pt, mlab, align=right]{decision cost\\(wall time, tokens)}
  (runner.west);
\draw[marr] (oracle.south) to[bend left=16]
  node[right=1pt, mlab, align=left]{declared cost,\\settled cost}
  (runner.east);
\draw[marr] ([xshift=20mm]runner.north west) to[bend left=14]
  node[right=3pt, mlab, pos=0.35]{BudgetView (read-only)}
  ([xshift=-4mm]driver.south east);
\end{tikzpicture}
\caption{Overview of the dual architecture. Solid arrows show the
query--result exchange between the Driver and Oracle; dashed arrows show accounting
and budget information mediated by the campaign runner. The runner charges both
parties against a common multi-resource budget and records each metered transaction
in an append-only ledger. The Driver receives a read-only \code{BudgetView}, enabling
budget-adaptive behavior without granting direct control over expenditure.}
\label{fig:arch}
\end{figure}

\cadaques{} separates the discovery loop into the two structural interfaces shown in
\cref{fig:arch}. Both are defined as Python \code{typing.Protocol} classes, so
conformance is structural: any object implementing the required methods can
participate without inheriting from a framework-specific base class or registering
with the package. This section presents the data model, the two protocols, the
campaign loop, and the budget and ledger semantics. Their implementation is described
in \cref{sec:impl}.

\subsection{Data model}
\label{sec:design:data}

The core query--response exchange is represented by three value types:
\code{Cost}, \code{Query}, and \code{Result}. They are implemented as frozen
dataclasses, which prevents their fields from being rebound after construction and
reflects their role as records of campaign events. Where fields contain mappings,
the immutability is shallow unless the supplied mappings are themselves immutable.
The runner, driver, and ledger derive their state from these records rather than from
separate mutable representations of the same event.

The first type is \code{Cost}, a frozen dataclass over the resource tuple
$(\code{seconds}, \code{cpu\_hours}, \code{euros}, \code{tokens})$. It supports
componentwise addition, subtraction and scaling, dictionary conversion, and a
componentwise dominance relation through \code{dominated\_by}. All components default
to zero, so \code{Cost(seconds=3.0)} denotes an expense of three seconds and
\code{Cost()} denotes zero recorded expenditure. An oracle or driver therefore needs to
report only the resource components it consumes.

The dominance relation reflects the fact that vector-valued resource costs are partially,
rather than totally, ordered. An expense is affordable only when each of its
components fits within the corresponding remaining budget. A scalar representation
is possible only after the user supplies an external valuation that makes the
resources commensurable. Keeping them as separate components therefore implements R1
without embedding a campaign-specific exchange rate in the stored records.

The second type is \code{Query}, which contains two mappings and an optional string
tag. The \code{params} mapping holds the design variables, while \code{fidelity}
holds parameters that control the cost and statistical or numerical quality of the
answer, such as lattice size, mesh density, or sampling effort. Plain mappings keep
the framework domain-agnostic: their keys belong to the application, and the core
package imposes no domain schema.

Representing fidelity separately from the design variables makes multi-fidelity
campaigns explicit at the type level rather than dependent on a naming convention
inside \code{params}. The same design point evaluated at two fidelities constitutes
two distinct queries, potentially with different declared costs, biases, variances,
or resolutions. The oracle's pricing and evaluation methods can therefore inspect
fidelity directly. The optional tag provides a lightweight bookkeeping channel, for
example to label phases of a scheduled policy. The evaluation in
\cref{sec:eval:fid} exercises this dimension empirically.

The third type is \code{Result}. It carries the query being answered, a scalar
\code{value}, a free-form \code{info} mapping, and the settled \code{Cost} of
producing the answer. Including the realized cost in the result ensures that the
driver receives the cost associated with every observation. Results reach the driver
both through the history passed to \code{propose} and through \code{observe}, so a
strategy may condition its future behavior on the expenditure associated with past
answers.

The \code{info} mapping provides a schema-free channel for domain-specific payloads,
such as convergence flags or diagnostic traces, without widening the core protocol.
Restricting \code{value} to a scalar is a deliberate scope decision of the present
release and is revisited in \cref{sec:discussion}.

\subsection{The Oracle and Driver protocols}
\label{sec:design:protocols}

\Cref{lst:protocols} shows the complete required method surface of the two protocols.
Optional metering hooks used by particular drivers are described separately below.

\begin{lstlisting}[float=t,caption={The required integration surface of \cadaques{}
(abridged docstrings). Conformance is structural (\code{typing.Protocol}).},
label={lst:protocols}]
@runtime_checkable
class Oracle(Protocol):
    def price(self, query: Query) -> Cost:
        """Declared (ex-ante) cost of answering."""
    def evaluate(self, query: Query) -> Result:
        """Answer; Result carries settled cost."""

@runtime_checkable
class Driver(Protocol):
    def propose(
        self,
        history: Sequence[Result],
        budget: BudgetView,
    ) -> Query:
        """Next query, given results and budget."""
    def observe(self, result: Result) -> None:
        """Feedback after each settled query."""
\end{lstlisting}

An \emph{Oracle} is any object that answers queries at a cost. Its \code{price}
method returns the estimated cost before the query is committed, allowing the runner
to test affordability. Its \code{evaluate} method performs the query and returns a
\code{Result} containing the realized cost. The two values need not agree and, in
physical or stochastic settings, often will not. Their discrepancy is retained as
campaign data, as described in \cref{sec:design:cost}.

A \emph{Driver} is any object that decides what to ask next. Its \code{propose}
method receives the complete sequence of settled results together with a read-only
\code{BudgetView}. The view exposes the total, spent, and remaining budget, together
with the largest fraction used across capped resource components. A strategy can
therefore condition its behavior on the resource state of the campaign, which
enables the budget-adaptive policies examined in \cref{sec:eval:rq1}. The view is
read-only: drivers can inspect the budget but cannot charge it directly. The
\code{observe} method delivers each newly settled result, supporting stateful
strategies such as incremental surrogate updates.

The two protocols have similarly small interfaces but distinct roles. Each is specified by
behavior rather than inheritance: simulators, instruments, and analytic models can
serve as Oracles, while random-search procedures, Bayesian optimizers, and LLM agents
can serve as Drivers. This structural separation allows implementations developed
outside \cadaques{} to participate through the same small contract.

\subsection{The campaign loop}
\label{sec:design:loop}

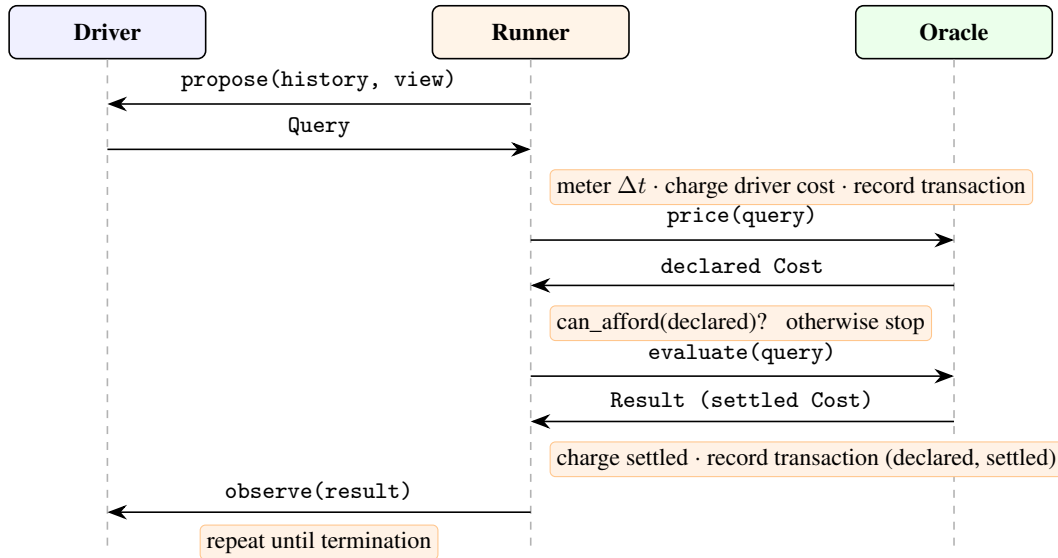
\begin{figure}[b]
\centering
\begin{tikzpicture}[font=\small, >={Stealth[length=2.6mm]},
  part/.style={draw, rounded corners=3pt, minimum width=26mm, minimum height=7mm,
               align=center, semithick, font=\small\bfseries},
  lifeline/.style={draw=gray!60, dashed, semithick},
  msg/.style={->, semithick},
  mlab/.style={font=\footnotesize\ttfamily},
  note/.style={font=\footnotesize, fill=orange!10, draw=orange!45, rounded corners=2pt,
               inner sep=2.5pt, align=center},
]
\node[part, fill=blue!6]   (D) at (0,0)    {Driver};
\node[part, fill=orange!9] (R) at (5.6,0)  {Runner};
\node[part, fill=green!8]  (O) at (11.2,0) {Oracle};
\draw[lifeline] (D.south) -- ++(0,-6.5);
\draw[lifeline] (R.south) -- ++(0,-6.5);
\draw[lifeline] (O.south) -- ++(0,-6.5);

\draw[msg] (5.6,-0.95) -- node[above=1pt, mlab]{propose(history, view)} (0,-0.95);
\draw[msg] (0,-1.55) -- node[above=1pt, mlab]{Query} (5.6,-1.55);
\node[note, anchor=west] at (5.85,-2.05)
  {meter $\Delta t$ $\cdot$ charge driver cost $\cdot$ record transaction};
\draw[msg] (5.6,-2.75) -- node[above=1pt, mlab]{price(query)} (11.2,-2.75);
\draw[msg] (11.2,-3.35) -- node[above=1pt, mlab]{declared Cost} (5.6,-3.35);
\node[note, anchor=west] at (5.85,-3.85)
  {can\_afford(declared)? \, otherwise stop};
\draw[msg] (5.6,-4.55) -- node[above=1pt, mlab]{evaluate(query)} (11.2,-4.55);
\draw[msg] (11.2,-5.15) -- node[above=1pt, mlab]{Result (settled Cost)} (5.6,-5.15);
\node[note, anchor=west] at (5.85,-5.65)
  {charge settled $\cdot$ record transaction (declared, settled)};
\draw[msg] (5.6,-6.35) -- node[above=1pt, mlab]{observe(result)} (0,-6.35);
\node[note] at (2.8,-6.75) {repeat until termination};
\end{tikzpicture}
\caption{Sequence of one campaign iteration. The runner first meters and records
the Driver's proposal, then requests the Oracle's declared cost and checks
affordability. If the query is admitted, the runner executes it, charges the settled
cost returned with the result, records the declared and settled values, and delivers
the result to the Driver.}
\label{fig:sequence}
\end{figure}

The \code{Campaign} object couples one Oracle to one Driver under one budget and owns
the loop summarized in \cref{fig:sequence}. Each iteration contains two metered
phases. In the \emph{decision phase}, the runner measures the elapsed duration of
the driver's \code{propose} call and records it as a settled driver transaction. A
driver may additionally expose a \code{last\_proposal\_cost} attribute to report
resource components that the runner cannot observe directly, such as LLM token use.
This is an optional metering convention rather than part of the required
\code{Driver} protocol. Implementations should avoid duplicating wall-time charges
already measured by the runner.

In the \emph{query phase}, the runner asks the Oracle for the declared cost and checks
whether that cost fits within the remaining capped resources. If it does, the runner
calls \code{evaluate}. The settled cost returned in the \code{Result} is then charged
unconditionally, because the realized cost is known only after the work has been
performed and may exceed the declaration. The driver and oracle transactions are
both appended to the ledger with their relevant declared and settled components,
implementing R3--R5. Because proposal cost is incurred before the Oracle's declared
cost is known, the final Driver transaction may be charged even when the resulting
query is rejected as unaffordable.

The loop terminates for one of three reasons. It stops with
\code{budget\_exhausted} when the next declared oracle cost is unaffordable, with
\code{driver\_stopped} when the driver raises \code{DriverStopped}, or with
\code{max\_queries} when an optional query-count safeguard is reached. Because
settlement is unconditional, a realized transaction may also bring a capped resource
to or beyond its limit before the next iteration. The returned
\code{CampaignResult} reports the stop reason together with the best result, complete
history, final budget state, and ledger.

The \code{CampaignResult.trace} method reconstructs a resource-normalized best-so-far
curve directly from the recorded transactions: the best observed value is expressed
as a function of cumulative settled expenditure in a selected resource component.
This illustrates the intended analysis pattern, in which resource-normalized artifacts
are derived from the campaign ledger rather than from separate bespoke logs.

\subsection{Budget semantics and the ledger}
\label{sec:design:cost}

A \code{Budget} may cap any subset of the supported resource components. Components
with positive totals are constrained, while the others are tracked without imposing
a limit. Affordability checks through \code{can\_afford} compare declared costs with
the remaining capped resources before commitment. Settlement through
\code{charge(..., settle=True)} is unconditional.

The read-only \code{BudgetView} supplied to drivers exposes \code{total},
\code{spent}, \code{remaining}, and \code{fraction\_used}. The last quantity is the
largest used fraction among capped resource components. For example, a campaign may
cap both \code{seconds=3600} and \code{tokens=100000}. A declared expense is
affordable only if both components fit within their respective remaining amounts.
The campaign becomes constrained as soon as any capped resource prevents
the next transaction. There is no framework-defined scalar total unless the user
provides an external valuation at analysis time.

The \code{Ledger} is an append-only sequence of frozen \code{Transaction} records.
Each record contains the responsible party (\code{oracle} or \code{driver}), a label,
the declared and settled costs, a timestamp, and free-form metadata. For oracle
transactions, the metadata includes the returned value and the query's design and
fidelity fields. The derived \code{overrun} property is the componentwise difference
between settled and declared costs and therefore quantifies transaction-level
cost-estimation error.

Systematic discrepancies between declared and settled costs can be detected and
analyzed from the ledger. They are directly available for post hoc calibration and
may support future cost-aware strategies if exposed through an appropriate adapter or
campaign interface. Aggregation methods such as \code{total} and \code{cumulative}
support resource analysis, while \code{to\_jsonl} exports one transaction per line in
JSON Lines format. This provides machine-readable transactional provenance suitable
for reproducible analysis and FAIR-oriented data workflows~\cite{Wilkinson2016}.

\section{Implementation}
\label{sec:impl}

The design of \cref{sec:design} is realized as a compact Python
package. This section describes the package organization, the reference Oracles and
Drivers included in the evaluated release, and the engineering and distribution
practices used to support testing, inspection, and reuse. The small implementation
surface is deliberate: it makes the campaign and accounting semantics easier to
inspect and audit directly.

\subsection{Package organization}
\label{sec:impl:organization}

\begin{figure}[t]
\centering
\begin{tikzpicture}[
  font=\small,
  mod/.style={draw, rounded corners=3pt, minimum height=12mm, minimum width=42mm,
              align=center, semithick, fill=orange!7},
  ext/.style={draw, rounded corners=3pt, minimum height=12mm, minimum width=48mm,
              align=center, semithick},
  pkg/.style={draw, rounded corners=4pt, inner sep=7pt, dashed, gray!70},
  dep/.style={-{Stealth[length=2.4mm]}, semithick},
  deplab/.style={font=\footnotesize, gray!40!black},
]
\node[mod] (cost)
  {\textbf{core.cost}\\[1pt]\footnotesize Cost, Budget, BudgetView};
\node[mod, right=20mm of cost] (protocols)
  {\textbf{core.protocols}\\[1pt]\footnotesize Query, Result, Oracle, Driver};
\node[mod, below=7mm of cost] (ledger)
  {\textbf{core.ledger}\\[1pt]\footnotesize Ledger, Transaction};
\node[mod, below=7mm of protocols] (campaign)
  {\textbf{core.campaign}\\[1pt]\footnotesize Campaign, CampaignResult};

\draw[dep] (protocols) -- (cost);
\draw[dep] (ledger) -- (cost);
\draw[dep] (campaign) -- (protocols);
\draw[dep] (campaign) -- (ledger);
\draw[dep] (campaign.north west) -- (cost.south east);

\begin{scope}[on background layer]
\node[pkg, fit=(cost)(protocols)(ledger)(campaign),
      label={[gray!70, font=\footnotesize\bfseries]above:cadaques.core}] (core) {};
\end{scope}

\node[ext, fill=blue!6, below=11mm of ledger.south, anchor=north, xshift=-4mm]
  (drivers)
  {\textbf{drivers}\\[1pt]\footnotesize RandomDriver, AnnealedLocalDriver};
\node[ext, fill=green!8, below=11mm of campaign.south, anchor=north, xshift=4mm]
  (oracles)
  {\textbf{oracles}\\[1pt]\footnotesize AnalyticOracle, Ising2DOracle};

\draw[dep] (drivers.north) --
  node[left=2pt, deplab]{imports public core types}
  (drivers.north |- core.south);
\draw[dep] (oracles.north) -- (oracles.north |- core.south);
\end{tikzpicture}
\caption{Package organization and import dependencies. The dashed boundary
encloses \code{cadaques.core}. Arrows point from a module to the core module on which
it depends. The reference \code{drivers} and \code{oracles} subpackages import only
public core types, illustrating the intended extension pattern for external
implementations.}
\label{fig:modules}
\end{figure}
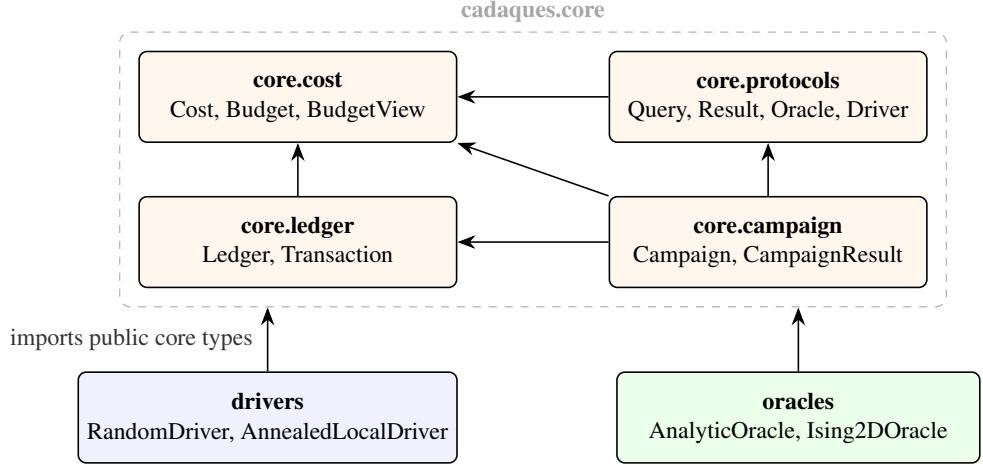

The package has three principal subpackages. Their import structure is shown
in \cref{fig:modules}, and their composition is summarized in
\cref{tab:modules}. The framework proper resides in \code{cadaques.core}, which
contains the data model and protocols (\code{protocols}), the cost and budget types
(\code{cost}), the campaign runner (\code{campaign}), and the transactional ledger
(\code{ledger}). Reference implementations reside in \code{cadaques.oracles} and
\code{cadaques.drivers}. They depend only on the public core types, illustrating the
intended extension pattern: domain packages can provide protocol-conformant Oracles
and Drivers without modifying the framework or importing its private implementation
details.

The top-level \code{cadaques} namespace re-exports fourteen public names:
\code{Budget}, \code{BudgetExceeded}, \code{BudgetView}, \code{Campaign},
\code{CampaignResult}, \code{Cost}, \code{Driver}, \code{DriverStopped},
\code{Ledger}, \code{Oracle}, \code{Query}, \code{Result},
\code{Transaction}, and \code{\_\_version\_\_}.

The v0.1.0 implementation contains 761 non-blank lines of Python, including
docstrings, and has one runtime dependency, NumPy~$\geq 1.24$. It requires
Python~$\geq 3.10$. Record types are implemented with frozen dataclasses, and the two
integration interfaces use \code{typing.Protocol} with
\code{@runtime\_checkable}. The limited dependency footprint is useful in discovery
workflows, which may run alongside instrument-control software or within tightly
managed computational environments.

\begin{table}[t]
\centering
\caption{Composition of \cadaques{} v0.1.0. LOC denotes non-blank Python lines,
including docstrings.}
\label{tab:modules}
\setlength{\tabcolsep}{7pt}
{\small
\begin{tabular}{@{}llr@{}}
\toprule
Module & Contents & LOC \\
\midrule
\code{core.protocols} & \code{Query}, \code{Result}, \code{Oracle}, \code{Driver} & 80 \\
\code{core.cost} & \code{Cost}, \code{Budget}, \code{BudgetView} & 138 \\
\code{core.campaign} & \code{Campaign}, \code{CampaignResult} & 146 \\
\code{core.ledger} & \code{Ledger}, \code{Transaction} & 92 \\
\code{oracles.analytic} & \code{AnalyticOracle}, test surfaces & 50 \\
\code{oracles.ising} & \code{Ising2DOracle} & 134 \\
\code{drivers.reference} & \code{RandomDriver}, \code{AnnealedLocalDriver} & 96 \\
\code{\_\_init\_\_} files & public API re-exports & 25 \\
\midrule
Total & & 761 \\
\bottomrule
\end{tabular}}
\end{table}

\subsection{Reference oracles}
\label{sec:impl:oracles}

The evaluated release includes two reference Oracles, intended respectively for
lightweight testing and controlled stochastic experiments.

\code{AnalyticOracle} wraps a function
$\code{fn}:\code{params}\mapsto\mathbb{R}$ as a metered Oracle. A configurable
pricing function supplies the declared cost and defaults to
\code{Cost(seconds=1.0)}. The settled cost combines the configured fixed charge with
the measured execution time of the wrapped function. This construction provides a
simple test case in which declared and settled costs
remain distinct even for inexpensive analytic objectives.

\code{Ising2DOracle} implements a checkerboard Metropolis Monte Carlo
simulation~\cite{Metropolis1953} of the two-dimensional Ising model. At a queried
temperature, it returns a finite-size, finite-sampling estimate of the magnetic
susceptibility. The implementation uses natural units, \(J=k_{\mathrm B}=1\), so
temperature is represented numerically in units of \(J/k_{\mathrm B}\). The exact
critical temperature of the infinite lattice in these units is~\cite{Onsager1944}
\[
\Tc=\frac{2}{\ln(1+\sqrt{2})}\approx 2.269
\]

The query fidelity is specified by the lattice size $L$, the number of
equilibration sweeps $n_{\mathrm{equil}}$, and the number of measurement
sweeps $n_{\mathrm{meas}}$. These parameters affect both computational
expense and estimator quality. The declared wall-time cost is obtained from a
calibrated model proportional to
\[
L^2\left(n_{\mathrm{equil}}+n_{\mathrm{meas}}\right),
\]
whereas the settled cost is the measured elapsed duration of the evaluation.
Declared and settled costs can therefore differ when the calibrated model does not
match the execution environment.

Each evaluation derives a separate pseudorandom stream from a common base seed. This
avoids reusing identical random sequences across repeated queries while preserving
the reproducibility of stochastic outputs under fixed software and numerical
conditions. Measured wall-time costs remain dependent on the execution environment
and are therefore not expected to reproduce bit-for-bit.

\subsection{Reference drivers}
\label{sec:impl:drivers}

The release includes two reference Drivers chosen to demonstrate the distinction
between budget-blind and budget-adaptive behavior.

\code{RandomDriver} samples each design variable uniformly within a box-bounded
domain at fixed fidelity. It does not use the observation history or the budget
state. Uniform random search provides a simple and important baseline for evaluating
more structured search procedures~\cite{Bergstra2012}.

\code{AnnealedLocalDriver} is a budget-adaptive local search procedure. It proposes
Gaussian perturbations around the point with the highest observed value. The
perturbation scale is interpolated linearly from \(\sigma_{\max}\) to
\(\sigma_{\min}\) as \code{BudgetView.fraction\_used} increases from zero to one.
The procedure therefore samples broadly early in the campaign and becomes
progressively more localized as the budget is consumed. It is included to
demonstrate that a Driver can condition directly on budget state, rather than as a
claim of algorithmic novelty or general superiority. Its role is examined in
\cref{sec:eval:rq1}.

\subsection{Engineering and distribution}
\label{sec:impl:engineering}

For v0.1.0, the repository contains a focused unit-test suite covering cost arithmetic,
affordability and settlement, ledger aggregation and export, campaign termination,
the read-only budget view, and reference-Oracle behavior. Continuous integration
runs the test suite on CPython 3.10--3.12. These tests provide regression protection
and encode several central semantics of \cref{sec:design:cost}, including
componentwise affordability, unconditional settlement, transaction recording, and
budget-governed termination.

Static type checking is performed with \code{mypy}, and source style is checked with
\code{ruff}. The package uses a \code{src} layout and is built with
\code{hatchling}. The core package has one runtime dependency, NumPy; the
reproduction scripts use additional scientific-Python packages where required.
Version 0.1.0 is distributed through PyPI (\code{pip install cadaques}). The
source code, documentation, tests, and reproduction scripts are available at
\url{https://github.com/jorgebravoabad/cadaques}, and the evaluated release is
archived at Zenodo under \doi{10.5281/zenodo.21293589}. The repository also
includes a \code{CITATION.cff} file containing machine-readable citation metadata.

\section{Evaluation}
\label{sec:eval}

The evaluation is designed to exercise the architectural claims of
\cref{sec:design,sec:impl} in operation rather than to establish the general
superiority of any search strategy. It combines a constructive demonstration of
expressibility, two controlled studies under common nominal wall-time budgets on a
task with an exact thermodynamic-limit reference, an empirical examination of declared
and settled costs, and a microbenchmark of the metering machinery. Results are
reported as observed, including outcomes unfavorable to the adaptive strategies,
because exposing such outcomes is one purpose of transaction-level accounting. The
evaluation addresses four research questions.

\begin{itemize}
\item[\textbf{RQ1}] (\emph{Expressibility}) Does exposing budget state through the
campaign interface enable strategies that a budget-blind interface supports only
through ad hoc external accounting, which would in addition omit the cost of
deliberation itself?

\item[\textbf{RQ2}] (\emph{Failure case under a common nominal wall-time budget})
How do planner deliberation and failure modes induced by noise affect three
protocol-conformant strategies under the same nominal 10\,s budget?

\item[\textbf{RQ3}] (\emph{Fidelity economics}) Under the same nominal 10\,s budget,
how do fixed low-fidelity, fixed high-fidelity, and budget-scheduled fidelity policies
compare?

\item[\textbf{RQ4}] (\emph{Overhead}) What runtime overhead does transaction-level
metering add to the campaign loop?
\end{itemize}

\subsection{RQ1: Expressibility}
\label{sec:eval:rq1}

We answer RQ1 constructively. The schedule of
\code{AnnealedLocalDriver} is a function of
\code{BudgetView.fraction\_used}: its perturbation scale depends on the fraction of
the campaign budget already consumed rather than on an iteration index. In a
budget-blind campaign interface, the same behavior would require external state or an
estimate of how many queries the available resources would support. That number is
not fixed in \cadaques{}, because query costs vary and driver deliberation is charged
against the same budget. A driver could in principle reconstruct budget state on its
own, by timing its calls and wrapping the oracle, but such ad hoc accounting would be
strategy-specific and would not necessarily be represented in the shared campaign
record. The claim is therefore one of standardized,
complete accounting rather than of strict expressive power. In the experiments of
\cref{sec:eval:rq2}, campaigns under the same nominal budget complete between 127
and 173 queries, illustrating the difference between iteration count and resource
expenditure.

The equal-budget comparisons in RQ2 are likewise implemented by the campaign runner
rather than reconstructed from fixed-length runs. The Gaussian process Driver used
there is defined in the accompanying reproduction script, imports only public
\cadaques{} types, and participates in the campaign without modifying the package.
Its proposal time is metered automatically by the runner. Together, these examples
demonstrate budget-dependent behavior and external protocol conformance through the
documented interface. In the representative systems reviewed in
\cref{sec:related}, we did not identify this combination of capabilities exposed through the
public campaign interface.

\subsection{RQ2: Failure case under a common nominal wall-time budget}
\label{sec:eval:rq2}

\subsubsection{Task}

We consider a discovery task based on the two-dimensional Ising model
($J = k_{\mathrm B} = 1$): locating its critical temperature by maximizing
the estimated magnetic susceptibility $\chi(T)$ over
$T \in [1.5, 3.5]$. The exact thermodynamic-limit value given in
Section~4.2 provides an external scoring reference. The Oracle, however,
returns a finite-size, finite-sampling susceptibility estimate. Its maximum
is noisy and may also be shifted from the critical temperature of the
infinite lattice, making the task a controlled but deliberately imperfect
proxy for stochastic discovery experiments.

\subsubsection{Setup}

Three Drivers are evaluated against \code{Ising2DOracle} at fixed fidelity
(\(L=24\), 400 measurement sweeps, and 200 equilibration sweeps). Each campaign has a
nominal wall-time budget of 10.0\,s on one CPU core. Experiments are repeated over
ten seeds, numbered 0--9, which determine both the Oracle's Monte Carlo streams and
the Drivers' stochastic choices.

The evaluated Drivers are the two reference implementations of
\cref{sec:impl:drivers} and a Gaussian process expected-improvement planner:
\code{RandomDriver}; \code{AnnealedLocalDriver}, with
\(\sigma_{\max}=0.5\) and \(\sigma_{\min}=0.02\), both expressed as fractions of the
width of the search interval; and GP-EI, using an RBF kernel with fixed hyperparameters, eight
initial random queries, and a 512-point candidate grid. The GP implementation is
defined outside the package as described under RQ1. Its surrogate update includes a
dense Cholesky factorization whose cost grows cubically with the history length in
this implementation, and its measured proposal time is charged by the runner.

\begin{figure}[t]
\centering
\includegraphics[width=0.92\textwidth]{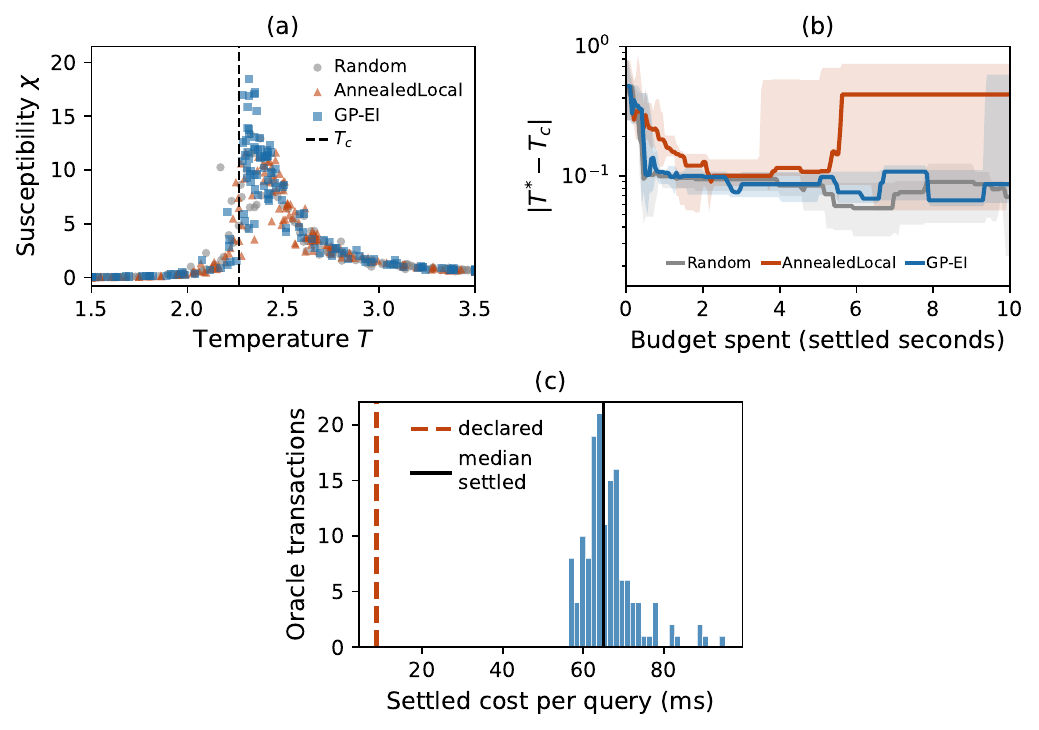}
\caption{Ising campaigns for RQ2 under a nominal 10\,s budget, over ten seeds.
(a)~Susceptibility observations versus queried temperature for seed 0; marker shape
identifies the Driver and the vertical dashed line marks the exact
\(\Tc\) of the infinite lattice.
(b)~Best-so-far temperature error \(|T^{*}-\Tc|\) versus cumulative settled
expenditure. Solid lines show medians across seeds and shaded bands show
interquartile ranges. The large upper quartiles of the two concentrating strategies
reflect sensitivity to incumbent capture.
(c)~Histogram of settled per-query costs for one GP-EI campaign, reconstructed from
its ledger. Vertical lines mark the Oracle's constant declared cost and the median
settled cost. On the evaluation machine, the median settled cost exceeds the
declared cost by \(652\%\), a factor of approximately 7.5.}
\label{fig:ising}
\end{figure}

Campaigns stop when the next declared query is unaffordable, completing 127--173
queries. The range reflects variation in settled query times and, for GP-EI, budget
spent on deliberation. Because settled costs are charged after execution and may
exceed their declarations, final expenditure can slightly exceed the nominal cap.
The quantitative panels in \cref{fig:ising,fig:fidelity} are reconstructed from
exported ledgers.

All timings in RQ2--RQ4 were measured on one core of a virtualized Intel Xeon
processor at 2.80\,GHz running Ubuntu 24.04, CPython 3.12.3, and NumPy 2.4.4.
Absolute timings are specific to this environment; their implications are discussed
in \cref{sec:eval:threats}.

\begin{table}[t]
\centering
\caption{Absolute error \(|T^{*}-\Tc|\), where \(T^{*}\) is the temperature with the
largest observed susceptibility, after campaigns with the same nominal 10\,s budget.
Results summarize ten seeds. The Fail column counts campaigns with error \(>0.3\).
Driver share is the fraction of total settled expenditure consumed by proposal
deliberation.}
\label{tab:results}
\setlength{\tabcolsep}{4pt}
{\small
\begin{tabular}{@{}lccccc@{}}
\toprule
Driver & Median & IQR & Max & Fail & Driver share \\
\midrule
Random & \textbf{0.069} & [0.024, 0.087] & 0.597 & 1/10 & 0.05\% \\
AnnealedLocal & 0.425 & [0.087, 0.733] & 0.769 & 5/10 & 0.10\% \\
GP-EI & 0.086 & [0.062, 0.605] & 0.769 & 3/10 & 3.2\% \\
\bottomrule
\end{tabular}}
\end{table}

\subsubsection{Results}

\Cref{tab:results,fig:ising} summarize the results. \Cref{fig:ising}a shows that all three Drivers sample the critical region in the representative seed, although their sampling patterns differ. Random search has the smallest
observed median error, 0.069, and one failure above the 0.3 threshold. The
budget-adaptive local strategy fails in five of ten campaigns. GP-EI spends 3.2\% of
its settled expenditure on deliberation and has a median error of 0.086.

A one-sided Mann--Whitney \(U\) test of the alternative that GP-EI yields smaller errors than Random gives
\(p=0.92\). No examined pairwise comparison reaches \(\alpha=0.05\) with ten seeds.
The data therefore do not support superiority of either concentrating strategy over
random search at this budget. More importantly here, the ledger captures the planner
cost and the resulting failure trajectories under the same resource constraint.

The dominant observed failure mode is \emph{incumbent capture}. A finite-sampling
susceptibility estimate can produce a spuriously large value far from \(\Tc\).
Strategies that subsequently concentrate around the best observation may then devote
further queries to that region. The annealed Driver localizes around its incumbent by
construction, while the incumbent that defines the GP-EI improvement threshold is set by the
largest observed value.
Five annealed and three GP-EI campaigns finish with errors between 0.6 and 0.77,
including outcomes near a domain boundary.
This failure mode is orthogonal to the budget mechanism itself: exposing budget
state makes schedules such as annealing expressible, but it does not protect a
strategy whose concentration rule is misled by noise. Budget-adaptivity is a
capability of the interface, not a guarantee of performance.

Random search does not adaptively concentrate around a spurious incumbent, although
it can still select a maximum produced by noise as its final estimate, as occurs in one
campaign. A fixed-\(N\) comparison would also omit the variation in query cost and,
unless measured separately, the GP's proposal cost. Economic termination and
symmetric charging therefore produce a resource-matched account of the three
strategies.

\subsubsection{Declared and settled costs}

\Cref{fig:ising}c compares the settled per-query costs of one GP-EI campaign with
the Oracle's declared estimate. The ex-ante cost model was calibrated on
different hardware and was deliberately left unchanged for this experiment. This
intentionally stale calibration provides a stress test of the distinction between
admission-time declarations and post-execution settlements.
On the evaluation machine, the median percentage overrun is computed as
\[
100\,
\frac{c_{\mathrm{settled}}-c_{\mathrm{declared}}}
     {c_{\mathrm{declared}}},
\]
where $c_{\mathrm{declared}}$ is the Oracle's predicted per-query cost and
$c_{\mathrm{settled}}$ is the measured cost recorded after execution. The
median overrun is 652\%, meaning that the median settled cost is approximately
7.5 times the declared value.

This discrepancy does not invalidate the comparisons, which are analyzed against
cumulative settled expenditure. It does affect affordability prediction: a planning
calculation based only on the declared model would substantially overestimate the
number of queries supported by the budget. The ledger exposes the calibration error
without a separate logging mechanism. Because declared and settled values are
recorded per transaction, the same ledger also provides the data needed to
recalibrate the cost model for the current environment; the reported overrun
therefore reflects a deliberately uncorrected model rather than a floor on the
calibration achievable with the recorded data.

The experiment is low-dimensional, noisy, and limited to one budget scale. The GP
implementation is also minimal, with fixed hyperparameters and no noise-robust
incumbent treatment. Its purpose here is narrower: planner cost, observed performance,
and cost-model error are recorded together, including when the planner performs
poorly.

\subsection{RQ3: Fidelity economics under a common nominal wall-time budget}
\label{sec:eval:fid}

RQ2 holds fidelity fixed. RQ3 varies it to exercise the representation of fidelity as
a metered query component (\cref{sec:design:data}). Three policies are evaluated over
the same ten seeds with nominal 10\,s budgets.

Two policies use uniform random sampling at fixed fidelity. The low-fidelity policy
uses \(L=16\) and 150 measurement sweeps, with a mean settled cost of approximately
40\,ms per query on the evaluation machine. Its estimates are inexpensive but noisier
and more affected by finite-size bias. The high-fidelity policy uses \(L=32\) and
800 measurement sweeps, at approximately 145\,ms per query, producing a less noisy
but more expensive estimate.

The third policy is a two-phase schedule implemented in the accompanying script.
During the first half of the budget it samples low-fidelity queries uniformly over
the full domain. During the second half it samples high-fidelity queries uniformly
within \(\pm0.25\) of the best low-fidelity temperature. The policy does not adapt
repeatedly around each new high-fidelity incumbent, and is therefore less exposed to
the specific feedback mechanism observed in RQ2, although its second phase can still
be misdirected by a misleading low-fidelity result.

Policies containing high-fidelity observations are scored using their high-fidelity
queries only, because susceptibility values are not directly comparable across
lattice sizes. The low-fidelity arm is necessarily scored from its own observations.
All arms are evaluated by the temperature error relative to \(\Tc\), but their
underlying estimators differ in finite-size bias and noise, as discussed in
\cref{sec:eval:threats}.

\begin{table}[t]
\centering
\caption{Fidelity policies under the same nominal 10\,s budget, over ten seeds. Fail
denotes error \(>0.3\). The two-phase policy is the only arm for which all ten
observed errors are below 0.1.}
\label{tab:fidelity}
\setlength{\tabcolsep}{3.6pt}
{\small
\begin{tabular}{@{}lccccc@{}}
\toprule
Arm & Median & IQR & Max & Fail & \shortstack{Mean settled\\cost/query} \\
\midrule
Random high-fi & 0.116 & [0.053, 0.548] & 0.685 & 4/10 & 144\,ms \\
Random low-fi & 0.068 & [0.030, 0.107] & 0.130 & 0/10 & 40\,ms \\
Two-phase & 0.067 & [0.040, 0.084] & \textbf{0.098} & 0/10 & 62\,ms \\
\bottomrule
\end{tabular}}
\end{table}

\begin{figure}[t]
\centering
\includegraphics[width=0.92\textwidth]{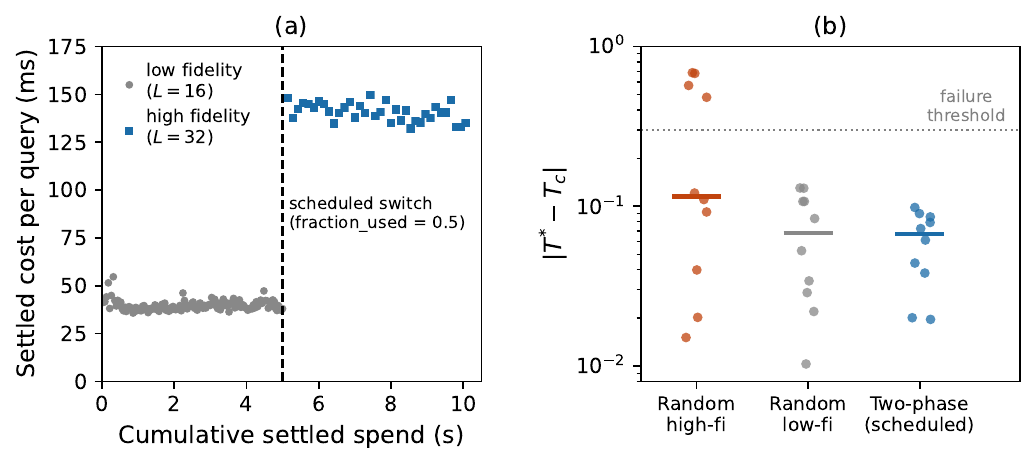}
\caption{Fidelity economics in RQ3.
(a)~Settled cost per query in one two-phase campaign as a function of cumulative
settled expenditure. The vertical dashed line marks the scheduled switch at
\code{fraction\_used}=0.5, from low-fidelity full-domain sampling to high-fidelity
sampling in the selected region. The phases consume approximately 5.0\,s each and
contain 126 and 36 queries, respectively.
(b)~Final temperature error for each policy under the same nominal budget. Each point
represents one seed, horizontal segments mark sample medians, and the dotted line
marks the error threshold of 0.3. The two-phase policy is the only arm for which all
ten observed errors are below 0.1.}
\label{fig:fidelity}
\end{figure}

\subsubsection{Results}

\Cref{tab:fidelity,fig:fidelity} report the comparison. In this experiment, the
two-phase policy has smaller errors than fixed high-fidelity sampling in a one-sided Mann--Whitney test (\(p=0.038\)). Its median is nearly identical to that of
the low-fidelity arm, 0.067 versus 0.068 (\(p=0.31\)), while its largest observed
error is smaller, 0.098 versus 0.130. With ten seeds and multiple examined contrasts,
these \(p\)-values are exploratory rather than confirmatory.

The fixed high-fidelity policy yields the least favorable aggregate results in this
experiment.
At the measured cost per query, its budget supports only 68--72 evaluations, reducing
domain coverage and the opportunity to average over occasional susceptibility
spikes. It exceeds the 0.3 failure threshold in four of ten campaigns.

At this budget scale, the extra fidelity of every query does not compensate for the
loss of coverage. The two-phase policy allocates inexpensive queries to reconnaissance
and more expensive queries to refinement, and produces the smallest observed
worst-case error. The comparison also shows that policies with markedly different
per-query costs can be run and evaluated within the same resource-accounted campaign.

\subsection{RQ4: Metering overhead}
\label{sec:eval:rq4}

We measured the end-to-end overhead of the campaign loop using 5{,}000-query runs of
\code{RandomDriver} against an \code{AnalyticOracle} wrapping an inexpensive
closed-form objective with zero declared cost. A bare Python loop performing the same
sampling and evaluation served as the reference.

Across three repetitions, the median bare-loop time was 4.0\,\textmu s per iteration,
whereas the fully metered campaign required 28.8\,\textmu s. The difference,
approximately 25\,\textmu s per query, includes proposal timing, budget updates,
ledger appends, and result bookkeeping.

An Oracle evaluation in RQ2 has a median settled cost of roughly
\(65\,\mathrm{ms}\) on the same machine. The measured framework overhead is
therefore more than three orders of magnitude smaller and corresponds to
approximately \(0.04\%\) of the Oracle evaluation cost in that setting.
For the workloads evaluated here, transaction-level accounting adds negligible
runtime relative to Oracle execution.

\subsection{Threats to validity}
\label{sec:eval:threats}

\paragraph{Construct validity.}
RQ2 scores a policy by the temperature associated with its largest observed
susceptibility. The finite-sampling estimator can generate isolated spikes, and the
study shows that these spikes affect both the adaptive trajectories and the final
scoring rule. A noise-robust evaluation procedure, such as repeated local
re-estimation around candidate maxima, could change the observed ordering. In
addition, the finite-\(L\) susceptibility maximum need not coincide exactly with the
\(\Tc\) of the infinite lattice.

In RQ3, susceptibility values are not directly comparable across lattice sizes.
Mixed-fidelity policies are therefore scored using high-fidelity observations, while
the low-fidelity arm is scored using its own estimator. The final temperature errors
share the same reference value, but the estimators differ in finite-size bias and
noise. For the two-dimensional Ising model, the finite-size shift of the
susceptibility maximum scales as \(L^{-1/\nu}\) with \(\nu=1\), of order
\(0.05\)--\(0.1\) for \(L=16\) and therefore comparable to the median errors
reported in \cref{tab:fidelity}. Part of the low-fidelity arm's temperature error is
thus expected to be systematic rather than statistical, which further complicates a
direct comparison of its median with that of the two-phase policy. The comparison
should therefore be interpreted as the quality of the answer
each policy can deliver within its budget, not as a controlled comparison of
identical estimators.

\paragraph{Internal validity.}
Ten seeds support only coarse statistical conclusions. RQ2 provides no evidence that
the adaptive strategies outperform random search, but it does not establish the
superiority of random search in general. The nominal RQ3 \(p=0.038\) contrast is
exploratory because of the small sample and multiple examined comparisons.

Settled costs are measured in wall time and are sensitive to machine load.
Seeds determine the stochastic streams but not the exact realized timings. The
stochastic query outputs are reproducible under fixed software and numerical
conditions, whereas budget-dependent trajectories may differ because measured wall
times affect when the campaign terminates. Publishing the ledgers makes this variation
observable but does not eliminate it.

\paragraph{External validity.}
The evaluation uses a single one-dimensional simulated task selected because its
target is known in the thermodynamic limit. The GP-EI implementation is deliberately minimal and uses
fixed hyperparameters. Behavior in higher-dimensional spaces, on real instruments,
under parallel or asynchronous execution, and with monetary or token budgets remains
to be demonstrated.

The overhead benchmark was conducted on one machine, one Python version, and one
inexpensive objective. Absolute timings will vary across environments. The large
observed separation between framework overhead and Oracle execution provides
headroom, but it should not be interpreted as a universal overhead guarantee.

\section{Related Work}
\label{sec:related}

\cadaques{} lies at the intersection of several overlapping bodies of work:
Bayesian optimization libraries, orchestration platforms for autonomous
experimentation, benchmarking frameworks for experiment planners, provenance and
workflow infrastructures, and scientific
agents based on large language models. These categories are used here for
organization rather than as mutually exclusive classes. For example, an optimization
library may also manage experiments, while a laboratory platform may include
planning and benchmarking components.

\Cref{tab:related} compares the architectural abstractions emphasized in the cited
publications and documented interfaces. It is not an inventory of every software
version or custom extension. The systems serve different roles: optimization libraries
provide decision algorithms, laboratory platforms coordinate physical operations,
benchmarking suites provide standardized tasks, and \cadaques{} supplies campaign-level
budget-governed admission and transaction records.

\begin{table}[t]
\centering
\caption{Qualitative positioning of representative systems relative to
\cadaques{}. The comparison concerns the cited publications and documented public
abstractions; it does not attempt to cover all functionality obtainable through
custom extensions or later software versions.}
\label{tab:related}
\setlength{\tabcolsep}{4.5pt}
{\small
\begin{tabularx}{\textwidth}{@{}p{24mm}p{38mm}X@{}}
\toprule
System & Primary emphasis & Relationship to the present work \\
\midrule
BoTorch/Ax
&
Bayesian optimization modeling and adaptive-experiment management
&
Provides advanced acquisition functions, including cost-aware and multi-fidelity
methods, together with experiment and trial management. A BoTorch- or Ax-based
planner can serve as a \cadaques{} Driver, while the campaign runner records declared
and settled query costs and meters the planner's own decision time. \\

Olympus
&
Benchmarking noisy optimization and experiment-planning algorithms
&
Provides standardized benchmark surfaces, experimentally derived datasets, and
interfaces for comparing planners. \cadaques{} can supply budget-governed execution
and transaction records from which resource-normalized benchmark curves are derived. \\

Hyperband/\hspace{0pt}BOHB
&
Budget allocation across fidelities in hyperparameter optimization
&
Treat a fixed resource budget as an allocation parameter internal to the search
algorithm. \cadaques{} instead places the budget in the campaign contract, with
per-transaction declared and settled costs and metered planner deliberation. \\

AiiDA
&
Workflow automation and data provenance in computational science
&
Automatically records the full data lineage of high-throughput workflows as a
queryable provenance graph. \cadaques{} records transactional cost provenance
(declared and settled expenditure per query and decision) rather than data lineage;
the two forms of record are complementary. \\

ChemOS
&
Orchestration of autonomous laboratory workflows
&
Coordinates laboratory equipment, data exchange, and experiment-planning components
at an operational level broader than the scope of \cadaques{}. The Oracle/Driver
boundary may be used within or alongside such a platform as a portable interface for
campaign-level accounting. \\

\cadaques{}
&
Budgeted campaign execution and transactional provenance
&
Focuses narrowly on vector-valued resource budgets, ex-ante and ex-post costs,
symmetric charging of planning and evaluation, and machine-readable transaction
records. \\
\bottomrule
\end{tabularx}}
\end{table}

\subsection{Bayesian optimization libraries}

Bayesian optimization provides a broad family of methods for selecting informative
queries when evaluations are expensive~\cite{Shahriari2016,Snoek2012}. Libraries
such as BoTorch and Ax provide modern probabilistic models, acquisition functions,
and experiment-management facilities~\cite{Balandat2020}. Their functionality
includes cost-aware acquisition utilities and continuous multi-fidelity
optimization~\cite{Kandasamy2017}. In the self-driving-laboratory setting, Atlas
provides a library of Gaussian-process planners tailored to laboratory campaigns,
including multi-fidelity and cost-constrained acquisition
strategies~\cite{Hickman2025}. As with BoTorch and Ax, such planners address the
decision algorithm itself; an Atlas-based strategy could likewise serve as a
\cadaques{} Driver, with the campaign runner supplying budget governance and
transaction records.

A related line of work in hyperparameter optimization treats evaluation budget,
rather than iteration count, as the fundamental unit of allocation.
Successive-halving methods and Hyperband distribute a fixed resource budget across
candidate configurations evaluated at increasing fidelities~\cite{Li2018}, BOHB
combines this allocation scheme with model-based proposals~\cite{Falkner2018}, and
libraries such as Syne Tune expose these and related multi-fidelity schedulers
behind a common tuning interface~\cite{Salinas2022}. The
two-phase schedule evaluated in \cref{sec:eval:fid} is structurally a coarse
relative of these methods. In these systems, however, the budget is an allocation
parameter internal to the search algorithm. They do not expose a campaign-level
contract that records declared and settled costs for each transaction, meters the
planner's own deliberation, or preserves vector-valued resource components for post
hoc analysis, which is the focus of the present work.

In this setting, cost is commonly introduced as an input to the optimization model
or acquisition rule. This is distinct from making resource expenditure a property of
every campaign transaction. In the documented public abstractions examined here, and subject
to the scoping stated at the start of this section, we did not find a single
campaign-level contract that simultaneously represents vector-valued resource caps,
records both declared and settled cost for each evaluation, and charges planner
deliberation against the same budget.

A BoTorch or Ax strategy can be adapted to the Driver protocol while retaining its
modeling and acquisition machinery. \cadaques{} then supplies budget views,
affordability checks, decision-time metering, settlement of realized query costs, and
transaction records.

\subsection{Autonomous-experimentation platforms}

Autonomous-laboratory platforms address the operational integration required to
execute closed-loop experiments. Systems such as ChemOS coordinate planning
algorithms, databases, instruments, scheduling, and data exchange across heterogeneous
laboratory components~\cite{Roch2018,Abolhasani2023}. These responsibilities extend
well beyond the scope of \cadaques{}, which does not provide instrument drivers,
workflow scheduling, safety interlocks, user interfaces, or laboratory data
management.

The distinction is primarily one of abstraction level, not competing capability.
\cadaques{} defines a small accounting boundary around the interaction between a
question-selecting component and a question-answering component. An orchestration
platform could expose this boundary internally, or a campaign could use it alongside
the platform, to associate each planning decision and experimental action with
declared and settled resource expenditure. Conversely, a simulation-only campaign
can use the same contract without adopting a full laboratory-orchestration stack.

\subsection{Benchmarking frameworks}

Benchmarking frameworks such as Olympus standardize noisy objective surfaces,
experimental datasets, and interfaces for comparing experiment-planning
algorithms~\cite{Hase2021}. Their principal contribution is to make planner
comparisons more realistic and reproducible than comparisons restricted to simple
synthetic test functions.

\cadaques{} addresses a different part of the benchmarking problem. It does not
provide a collection of benchmark datasets or claim to prescribe which planners
should be compared. Instead, it supplies execution and provenance semantics through
which a benchmark can be run under explicit resource constraints. Evaluation-count
curves, wall-time curves, monetary-cost curves, and other resource-normalized
summaries can then be reconstructed from the same ledger. Olympus tasks could
therefore be exposed through an Oracle adapter, while Olympus-compatible planners
could be exposed through Driver adapters.

\subsection{Provenance and workflow systems}

Recording what a computational campaign did is also the concern of workflow and
provenance infrastructures. In computational materials science, AiiDA automatically
captures the full data provenance of high-throughput workflows as a queryable graph
of calculations and data, with reproducibility as its central goal~\cite{Huber2020}.
Experiment-tracking tools common in machine learning likewise log metrics,
parameters, and artifacts produced during runs. These systems answer the question of
\emph{what was executed and from which inputs}. \cadaques{} records a complementary
and narrower kind of provenance: the economic history of a campaign, in which each
query and each planning decision carries a declared and a settled resource cost and
is admitted against explicit budget caps. Data lineage is outside its scope, and
transactional cost accounting is outside the documented scope of the provenance
systems above; a campaign could use both, with a workflow engine tracking lineage
while the Oracle/Driver boundary tracks expenditure.

\subsection{LLM agents for science}

Recent systems use large language models to plan scientific procedures, select
tools, interpret observations, and interact with laboratory or computational
resources~\cite{Boiko2023,Bran2024}. Such planners introduce resource components that
are less visible in conventional optimization loops. Their decisions may consume
tokens, wall time, external API charges, and additional computational services.
Depending on the relative costs of planning and experimentation, omitting these
components may materially change the apparent efficiency of a campaign.

The architectural issue is not specific to language models. It is the more general
question of whether the cost of deciding what to evaluate is recorded alongside the
cost of performing the evaluation. \cadaques{} provides one explicit implementation
of this accounting principle by allowing Driver and Oracle transactions to draw from
the same vector-valued budget. A future LLM adapter could report token or monetary
charges through the optional proposal-cost hook described in
\cref{sec:design:loop}. The present release does not include or evaluate an
end-to-end LLM Driver.

\subsection{Positioning of \cadaques{}}

Cost-aware acquisition, multi-fidelity modeling, experiment orchestration,
provenance recording, and resource-constrained optimization are established ideas.
\cadaques{} combines four of them in a small campaign abstraction: typed resource
vectors, ex-ante declarations and ex-post settlements, common charging of query and
decision costs, and an append-only record for resource-normalized analysis. It is
designed to interoperate with existing optimization, benchmarking, and orchestration
software.

\section{Discussion and Future Work}
\label{sec:discussion}

This section discusses the practical implications, current limitations, design
choices, and next steps.

\subsection{Implications}

The evaluation points to three practical lessons. First, planner cost should be
considered together with sample efficiency. In the Ising study, random search remained
a competitive resource-normalized baseline, while the concentrating strategies were
vulnerable to noise-induced incumbent capture. Additional planner sophistication did
not guarantee better performance under the common budget.

Second, allocating resources across fidelities may matter as much as the choice of
search algorithm. In RQ3, inexpensive reconnaissance followed by higher-fidelity
refinement produced a smaller observed worst-case error than applying the
high-fidelity estimator throughout. At this budget, balancing coverage and fidelity
mattered more than maximizing the accuracy of every query.

Third, iteration-indexed performance curves are insufficient when evaluations and
decisions consume heterogeneous resources. They can obscure differences in query
cost, planner deliberation, fidelity, and final expenditure. Reporting performance
against cumulative settled resource use, together with the underlying transaction
record, makes these factors visible and permits alternative valuations to be applied
after the campaign. For LLM-based scientific agents, this distinction is especially
relevant: evidence that an agent is resource-efficient requires accounting for token,
time, and monetary costs as well as experimental outcomes.

\subsection{Limitations}

Version 0.1.0 supports sequential campaign execution only:
batch proposals, asynchronous evaluations, parallel instruments, and distributed
simulation workloads are not yet supported. Adding these modes will require explicit
semantics for reservations, concurrent settlements, cancellation, and ordering in
the ledger.

The current \code{Result} type contains a scalar objective value. Multi-objective,
constrained, vector-valued, and structured-output campaigns would require extensions
to the result representation and to the framework's generic notion of an incumbent.
The present implementation also assumes that a completed evaluation returns one
settled result. Partial observations, streaming measurements, failed evaluations,
and recoverable instrument states are outside the current model. Supporting
asynchronous execution and structured results will therefore require changes to the
data and execution models, not only additional adapters.

Affordability is checked using declared cost, whereas settled cost is known only
after execution and is charged unconditionally. Consequently, a transaction may
bring expenditure beyond a nominal cap when the declared model underestimates the
realized cost. This behavior reflects the irreversibility of completed work, but the
release does not yet provide reservations, uncertainty margins, probabilistic
affordability checks, or policies for handling large overruns.

Driver wall time is measured automatically within the runner. Other resource
components are self-reported through \code{last\_proposal\_cost}. This mechanism does
not independently verify token counts, remote API charges, GPU use, energy
consumption, or work performed asynchronously outside the measured call. Automatic
meters integrated with API clients, schedulers, and hardware telemetry are therefore
an important extension.

Finally, the empirical evaluation uses one simulated, one-dimensional discovery task
and does not demonstrate operation with a physical instrument, an end-to-end LLM
agent, or a genuinely multi-resource campaign. The study validates the implemented
accounting behavior and illustrates the analyses it enables, but it does not establish
performance or usability across the broader range of settings in autonomous discovery.

\subsection{Design alternatives}

Three principal design choices merit discussion. First, structural protocols were
chosen instead of abstract base classes. This allows objects defined in external
packages or user scripts to participate without inheriting from framework-owned
classes. Static type checkers can detect many interface mismatches during development,
and \code{@runtime\_checkable} permits limited runtime conformance checks. These
mechanisms do not verify method semantics or all type details at runtime, so some
errors may still emerge only when a campaign invokes the object. The trade-off favors
low coupling and third-party interoperability over strict runtime enforcement.

Second, costs are stored as vectors rather than collapsed into a scalar through
framework-defined exchange rates. A scalar budget would simplify ranking and
termination, but it would embed a valuation that depends on the campaign, facility,
deadline, and scientific objective. Preserving resource components separately keeps
the ledger reusable: users may apply different monetary or utility valuations to the
same transaction history without rewriting the original record. A related choice is
the fixed resource vocabulary of the present release: \code{Cost} spans exactly four
components (seconds, CPU hours, monetary cost, and tokens), chosen to cover the
computational and API-metered campaigns targeted by v0.1.0 while keeping the
arithmetic, dominance, and serialization semantics simple to specify and test.
Resources named in \cref{sec:background}, such as GPU hours, energy, instrument
time, or consumables, do not yet have dedicated components and would currently have
to be mapped onto the existing ones. Making the component set user-extensible
without weakening the componentwise budget semantics or fragmenting ledger schemas
across campaigns is a planned revision of the data model.

Third, metering responsibility is placed primarily in the campaign runner rather than
delegated entirely to Drivers and Oracles. The runner can measure elapsed time around
calls and can enforce a common accounting path for both parties. This measured wall
time is a lower-level observation that does not depend on voluntary reporting by the
component. It is not, however, a complete measure of all resource use. The optional
proposal-cost hook complements it by allowing adapters to report externally metered
components such as tokens or API charges. Future integrations should replace or
validate self-reported values wherever authoritative meters are available.

\subsection{Future work}

Near-term engineering priorities include campaign replay from exported ledgers,
checkpointing and resumption, batch proposals, asynchronous evaluation, and
concurrency-safe budget and ledger semantics. A reference Bayesian optimization
Driver would provide a reusable integration with an established Gaussian process
stack while allowing model-fitting and acquisition costs to be recorded through the
same campaign machinery.

A second priority is an end-to-end LLM Driver adapter with explicit token, monetary,
and wall-time metering. Such an adapter would enable controlled comparisons between
LLM-based planning, conventional optimization, and inexpensive baselines under common
resource constraints. Automatic accounting should be obtained from the relevant API
clients where possible rather than relying solely on self-reported costs.

On the Oracle side, protocol-conformant adapters for established simulation packages
and laboratory systems would test whether the abstraction remains practical when
evaluations involve job queues, failures, instrument availability, and heterogeneous
metadata. Photonics simulation is one candidate application, but meaningful
validation will require examples from multiple scientific domains.

Longer term, resource-normalized evidence should become easier to produce and compare
across autonomous-discovery studies. That will require agreed resource definitions,
reporting conventions for declared and settled costs, and benchmark protocols that
separate scientific performance from resource efficiency. \cadaques{} provides one
piece of that infrastructure.

\section{Conclusion}
\label{sec:conclusion}

We have presented \cadaques{}, a compact framework for autonomous discovery built
around the premise that resource cost should be represented within the discovery
loop rather than added only during post hoc analysis. Two small, role-specific
protocols separate the component that proposes queries from the component that
answers them. Around this boundary, the framework provides vector-valued budgets,
separate declared and settled costs, metering of both planning and evaluation, and an
append-only transaction ledger. Together, these mechanisms allow campaigns to run
under explicit resource control and performance to be reconstructed as a function of
expenditure from a single machine-readable record.

On the noisy Ising task, neither the budget-adaptive local strategy nor the Gaussian
process planner improved on random search under the same nominal wall-time budget in
our experiments.
Both concentrating strategies were vulnerable to noise-induced incumbent capture, and
the Gaussian process planner consumed 3.2\% of its settled expenditure in
deliberation. In the fidelity study, at the examined budget scale, a two-phase policy
that used inexpensive queries for exploration and more expensive queries for refinement
produced the smallest observed worst-case error. Evaluation counts alone would hide both the planner cost and
the fidelity trade-off.

The additional accounting machinery required approximately
\(25\,\text{\textmu s}\) per query in the reported microbenchmark, about three orders
of magnitude less than the Oracle runtime in the Ising experiments. As scientific
planners increasingly rely on computationally expensive surrogate models, external
services, or LLM agents, the cost of deciding what to evaluate may become a
non-negligible part of the campaign itself. \cadaques{} provides a small,
interoperable foundation for recording that cost alongside scientific outcomes and
for evaluating discovery strategies under explicit resource constraints.


\renewcommand{\bibfont}{\small}
\bibliographystyle{unsrtnat}
\bibliography{cadaques}

\end{document}